\documentclass[journal,graphicx]{IEEEtran}
\usepackage{graphicx}

\newcommand{\frametext}[1]
{
\noindent
\refstepcounter{figure}
\fbox
{
\parbox[t]{8.cm}
{#1 } }

\vspace{0.2cm} }

\usepackage{amsmath}

\usepackage{amssymb}
\usepackage[all]{xy}

\newcommand{\remove}[1]{}

\newtheorem{definition}{Definition}

\newtheorem{lemma}{Lemma}

\newtheorem{theorem}{Theorem}
\newtheorem{corollary}{Corollary}
\newenvironment{myitemize}
{\begin{list}{$\bullet$}{
\itemindent=-.3cm
\listparindent=.3cm 
\itemsep=0.0in
\parsep=0.0in
\topsep=0.0in
\partopsep=0.0in}}{\end{list}}

\newcounter{itemcount}

\begin{document}

\title{On the Optimality of    Keyless Authentication in a Noisy Model}
\author{Shaoquan Jiang
\thanks{Shaoquan Jiang is  with the Institute of Information Security,
Mianyang Normal University,
 Mianyang, China 621000.
Email: shaoquan.jiang@gmail.com} }
\maketitle

\begin{abstract}
We further study     the keyless authentication problem in a noisy model in our previous work, where no secret setup is available for sender Alice and receiver Bob while there is DMC $W_1$ from Alice to Bob and a two-way noiseless but insecure  channel between them. We propose a construction such that the message length over DMC $W_1$ does not depend on the size of the  source space.   If the source  space is ${\cal S}$ and the number of channel $W_1$ uses is $n$, then our protocol only has a round complexity of $\log^*|{\cal S}|-\log^*n+4.$ In addition, we show that  the round complexity of any secure protocol in our model is  lower bounded by  $\log^*|{\cal S}|-\log^* n-5$. We also obtain a lower bound on the success probability when the message size on DMC $W_1$ is given. Finally, we derive  the capacity for a non-interactive authentication protocol under general DMCs, which extends the result under  BSCs  in  our previous work.

\end{abstract}


\section{Introduction}
Message authentication is a protocol that allows a sender Alice to send a source state $S$ to a receiver Bob such that the latter is assured of the authenticity.  This mechanism was first studied  by \cite{GMS74} in a form of  a non-interactive protocol, called  a message authentication code (MAC).

Security of an  information system  usually is  quantified through analyzing   a number of attacks. There are two types of attacks for a message authentication protocol. In the  type I attack, the attacker Oscar plays between Alice and Bob and can modify, block, delete the messages over the channel. He succeeds if Bob finally accepts a source state $S'$ that is not authenticated by Alice. This is  known as a {\em substitution attack}. In the type II attack, Oscar impersonates Alice to directly authenticate a source state $S$ to Bob. He succeeds if Bob finally accepts $S.$ This is known as an {\em impersonation attack. }

The success probability of Oscar is closely related to his time complexity. A probabilistic polynomial time is a  widely adopted complexity class.  However, in this work, we are interested in the information theoretical security, where Oscar has an infinite time complexity. The advantage of this type of system  is that the security does not rely on  any hardness assumption (such as factoring assumption \cite{RSA78}).

To achieve authentication, Alice must have some resource that can distinguish herself from Oscar. For example, if Alice and Bob share a common secret \cite{GMS74}, then this secret can  play this role. In the literature, a signing key of a signature \cite{RSA78} and  a private key \cite{BCK98} of a public key encryption scheme are also examples of  this role.
In this work, we consider the case, where a noisy  channel for   Alice better in some sense than that  for Oscar will play as this role. 

Channel noise traditionally plays an undesired  role in many areas.  However,  Wyner \cite{Wyner75} showed that the channel noise  can be used to  establish a common secret for Alice and Bob.  Csisz\'{a}r and K\"{o}rner \cite{CK78} generalized this result  to a broadcast channel. Since then,  key agreement over a noisy channel has been extensively studied \cite{AC93,CN00,KDW08,HS11,Mau93,Mau03a,BTV12}. Other secure  mechanisms over a noisy channel were also studied; see \cite{CK88,CMW04,NW06} for oblivious transfers and \cite{BINS06,BBM07,Crea97,WNI03} for commitments.  Surveys on   information theoretical security over noisy channels can be found in \cite{LPS08,BB11}.

\subsection{Related works}
Authentication that uses a noise as an advantageous resource has been  studied in the literature but far from being well-studied (to our knowledge).
Baracca et al  \cite{BLT12} studied the physical layer authentication over MIMO  fading wiretap channels, where  they  assumed no shared key but an authenticated initialization from the sender to  the receiver.    Korzhik et al \cite{KY+07} considered an  authentication problem over a (noiseless) public discussion channel and   an  initialization using    noisy channels.   Lai et al  \cite{LGP09} considered a noisy authentication model with  a shared  key, where  the sender-receiver channel is better than the sender-adversary channel. Our previous work \cite{Jiang14} studied a new authentication model. In this model,  Alice  and Bob  share no key. There is a discrete memoryless  channel $W_1$ from Alice to Bob and a DMC  $W_2$ from Oscar to Bob. There is also a noiseless channel between any two of Alice, Bob and Oscar. Oscar can read any message  from Alice (over channel $W_1$ or noiselessly) or from  Bob (noiselessly)  in the clear text. He can also arbitrarily modify the message over the noiseless channel between Alice and Bob. But the message over channel $W_1$ can not be tampered. In addition, Oscar can impersonate Alice  to send any message to Bob using his channel $W_2$.
A characterization of the (in)existence in this model was given in \cite{Jiang14}. Given the existence,  an efficient construction was proposed. Further, the non-interactive authentication capacity with  BSC $W_1$ and BSC $W_2$  was given.  Authentication that tries to remove the noise pollution  on  the data  has been studied in the literature. For instance,  Martinian et al  \cite{MWC05} considered an  authentication with  a legal  distortion and  Yu et al \cite{YBS08} considered a covert authentication over a noisy channel. This type of work is not our  interest as we consider a noise as an advantageous resource to achieve the  authentication.

\subsection{Contribution}
This paper further studies      the keyless authentication problem in the  noisy model \cite{Jiang14}. We extend  the  construction in \cite{Jiang14} to authenticate  a source state  of any length using a fixed length $n$ of DMC messages over $W_1$, while in \cite{Jiang14}, $n$ heavily depends  on the size of the source space ${\cal S}$.   Our price is a round complexity of $\log^*|{\cal S}|-\log^*n+4$ while the protocol in \cite{Jiang14} has only 3 rounds. However, we show that the round complexity of any secure protocol in our model must be lower bounded by  $\log^*|{\cal S}|-\log^* n-5$. This shows that our protocol is nearly round optimal. We remark that this lower bound does not contradict the 3-round protocol in \cite{Jiang14} as $n\ge \frac{\log\log |{\cal S}|}{C}$ there, where $C$ is the Shannon capacity of $W_1$.  We also obtain a lower bound on the success probability of Oscar. Finally, we obtain  the capacity for a non-interactive authentication protocol in our model with general DMCs $W_1, W_2$ (which extends  of the result in \cite{Jiang14} with  BSC $W_1$ and BSC $W_2$), where the authentication capacity is   the maximum  achievable ratio $\frac{\log|{\cal S}|}{n}$.

\section{Preliminaries} \label{sect: symbol}
\noindent {\bf Notions. } \quad We list notions that will be used later.
\begin{itemize}
\item Random variable  is abbreviated as RV.
\item Denote a RV by a capital letter (e.g., $X$), its realization by a lower case letter (e.g., $x$) and  its alphabet space  by a calligraphic letter (e.g., ${\cal X}$).
\item $x^n$ denotes a sequence $x_1, \cdots, x_n$  of length $n$.
\item $P_X$ is  the distribution of $X$ (i.e.,  $P_X(x)=P(X=x)$). Similarly, $P_{Y|X}(b|a)\stackrel{def}{=}P(Y=b|X=a).$
\item $T_{z^n}(\cdot)$ for $z^n\in {\cal Z}^n$ is a distribution over ${\cal Z}$ with $T_{z^n}(u)$ being the fraction of  $u$ in $z^n$ for any $u\in {\cal Z}.$
\item $P_X^n(x^n)\stackrel{def}{=}\prod_{i=1}^n P_X(x_i)$.
\item i.i.d. denotes an  independent and  identical distribution.
    \item Function $negl(n)$ is {\em negligible} in $n$  if  for any polynomial $f(n)$, $\lim_{n\rightarrow \infty} negl(n) f(n)=0$.
\item $\log^{(j)} x=\underbrace{\log\cdots \log}_{j}(x)$ (i.e., the composition of $\log$ function for $j$ times).
\item $\log^*n$ is the minimum $i$ such that $\log^{(i)}n< 2.$
\item {Convex hull} $Cov(S)$ for  a set $S$ of vectors  is the set of all possible convex combinations of vectors in $S$.
\item $[n]$ denote the set $\{1, \cdots, n\}$.
\item For $S\subseteq [p]$ and a matrix  $W=(W_1,\cdots, W_p)^T$ with row vectors $W_1, \cdots, W_p$, define   $W_S=\{W_s\mid s\in S\}$.
\item {Statistical distance} between  RVs $X$ and $X'$ is  $\Delta(X,  X')=\sum_{x} |P_X(x)-P_{X'}(x)|.$  We also denote it  by  $\Delta(P_{X}, P_{X'})$. For any distribution $P$ over ${\cal X}$ and a {\em compact}   set of distributions ${\cal S}$ over ${\cal X}$, define  $\Delta(P, {\cal S}){=}\min_{Q\in {\cal S}} \Delta(P, Q).$
\item Hamming distance $d_H(x^n, y^n){=}|\{i\mid x_i\ne y_i, i\in [n]\}|.$
\item The binary entropy function $h(\alpha)=-\alpha\log\alpha-(1-\alpha)\log(1-\alpha)$ for $ \alpha\in [0,1].$

\end{itemize}
\subsection{Discrete memoryless channel}
A discrete channel with input $X$ over  ${\cal X}=\{a_1, \cdots, a_p\}$ and output $Y$ over  ${\cal Y}=\{b_1, \cdots, b_q\}$ is denoted by  a
stochastic matrix  $$W=\left(
\begin{array}{ccc}
W(b_1|a_1)& \cdots & W(b_q|a_1) \\
\vdots & \ddots& \vdots\\
W(b_1|a_p)& \cdots & W(b_q|a_p)
\end{array}
\right),
$$ where $W(y|x)=P_{Y|X}(y|x)$. In this case, we say $X$ and $Y$ are {\em connected by channel} $W$. The channel is {\em discrete memoryless} (DMC) if $P_{Y^n|X^n}(y^n|x^n)=\prod_{i=1}^n W(y_i|x_i).$   It is  {\em non-redundant} if $\Delta\Big{(}W_i, Cov(W_{[p]\backslash{\{i\}}})\Big{)}>0$ for any $i\in [p].$

{\em  A $n$-length code} ${\cal C}$ for $W: {\cal X}\rightarrow {\cal Y}$ with source  ${\cal S}$ is described by an encoding scheme  $f:  {\cal S}\rightarrow {\cal X}^n$ and a decoding scheme $\phi: {\cal Y}^n\rightarrow {\cal S}\cup\{\perp\}$. A decoding result  $\perp$ denotes a detection of  error. For $S\in {\cal S}$, $f(S)\in {\cal X}^n$ is called a {\em codeword}. When $f(S)$  is sent over $W$ and  received as  $Y^n\in {\cal Y}^n$, the receiver will decode it to $\phi(Y^n).$ If $\phi(Y^n)\ne S,$ an error occurs.     The {\em error probability}
is  $P(\phi(Y^n)\ne S)$.

\subsection{Typical sequences} \label{sect: inf}

In this subsection, we introduce the notions of typical and conditional typical sequences \cite{CK11}.
\begin{definition} Let $X$ be a RV over ${\cal X}$.  We say that $x^n\in {\cal X}^n$ is    {\em $\epsilon$-typical} if $|T_{x^n}(a)- P_X(a)|\le \frac{\epsilon}{|{\cal X}|}$ for any $a\in {\cal X}$ and whenever  $P_X(a)=0$, it holds that  $T_{x^n}(a)=0$.  The set of $\epsilon$-typical sequences for $X$ is denoted by $\textsf{T}_{[X]_\epsilon}^n$.
\end{definition}

\begin{definition} Let $X$ and $Y$ be RVs over  ${\cal X}$ and ${\cal Y}$ respectively. $y^n\in {\cal Y}^n$ is {\em conditionally $\epsilon$-typical} given    $x^n\in {\cal X}^n$,  if $|T_{x^ny^n}(a, b)-T_{x^n}(a)P_{Y|X}(b|a)|\le \frac{\epsilon}{|{\cal X}|\cdot|{\cal Y}|}$ for all $a\in {\cal X}, b\in {\cal Y}$ and whenever  $P_{XY}(a, b)=0$, it holds that $T_{x^ny^n}(a, b)=0$.
The set of conditionally $\epsilon$-typical sequences for $Y$, given $x^n$, is denoted by $\textsf{T}_{[Y|X]_\epsilon}^n(x^n)$, and also by    $\textsf{T}_{[W]_\epsilon}^n(x^n)$ if $X$ and $Y$ are connected by DMC $W$.
\end{definition}

 The following is a basic property  of typical sequences. The proof can be found in \cite[Chapter 2]{CK11}.
\begin{lemma} 
Let $X$ and $Y$ be RVs over ${\cal X}$ and ${\cal Y}$ respectively.
Then,
there exists  constants $\lambda_1>0$ and $\lambda_2>0$ such that
\begin{align*}
 P_Y^n(\textsf{T}^n_{[Y]_\epsilon})\ge& 1-2^{-n\lambda_1\epsilon^2}\\
 P_{Y|X}^n(\textsf{T}^n_{[Y|X]_\epsilon}(x^n)|x^n)\ge& 1-2^{-n\lambda_2\epsilon^2},\quad \forall x^n\in \textsf{T}_{[X]_\epsilon}^n,
\end{align*}
 when $n$ large enough.
\remove{
\begin{itemize}

\item[1.] For  any type $Q$ of ${\cal X}^n$,
\begin{equation}
(n+1)^{-|{\cal X}|}\cdot 2^{nH(Q)}\le |\textsf{T}_Q^n|\le 2^{nH(Q)}.
\end{equation}
\item[2.] If $\Delta(X_1; X_2)\le 1/2$, then
\begin{equation}
|H(X_1)-H(X_2)|\le \Delta(X_1; X_2)\log\frac{|{\cal X}|}{\Delta(X_1; X_2)}.
\end{equation}
\item[3.] There are at most $(n+1)^{|{\cal X}|}$ different types in ${\cal X}^n. $
\item[4.]  There exists  constants $\lambda_1>0$ and $\lambda_2>0$ such that when $n$ large enough, for any $x^n\in \textsf{T}_{[X]_\epsilon}^n$
\begin{eqnarray*}
 P_Y^n(\textsf{T}^n_{[Y]_\epsilon})&\ge& 1-2^{-n\lambda_1\epsilon^2}\\
 P_{Y|X}^n(\textsf{T}^n_{[Y|X]_\epsilon}(x^n)|x^n)&\ge& 1-2^{-n\lambda_2\epsilon^2}
\end{eqnarray*}
\end{itemize}
}
\label{le: basicH}
\end{lemma}

\subsection{Basic inequalities}

The following lemma is from \cite{Jiang14}. It essentially states that if the distribution $T_{Z^n}$ induced by the output  $Z^n$ of a DMC $W$ is  close to a distribution $P$, then $P$ must be close to $Cov(W)$.
\begin{lemma} Let  $P$ be a distribution over ${\cal Z}$.  Let  $Z^n$  be an output of DMC $W: {\cal X}\rightarrow {\cal Z}$ with input $X^n$.   If
\begin{eqnarray}
P_{Z^n}\Big{(}|T_{Z^n}(u)-P(u)|\le {\epsilon_1}, \mbox{ for all }  u\in {\cal Z}\Big{)}> \epsilon_2, \label{eq: Dbound-1}
 \end{eqnarray}
 for  some $\epsilon_1, \epsilon_2>0$,  then
 \begin{eqnarray}
 \Delta\Big{(}P; Cov(W)\Big{)}\le |{\cal Z}|\epsilon_1+|{\cal Z}|\sqrt{\frac{\ln{(2/\epsilon_2)}}{2n}}. \label{eq: Dbound}
\end{eqnarray} \label{le: neg}
\end{lemma}
The next lemma is  taken from \cite{Jiang14a}. It essentially states that if $x^n$ and $\bar{x}^n$ has a large distance, then sending $x^n$ through a non-redundant DMC $W$ is unlikely to result in an output $Y^n$ that is conditionally $\epsilon$-typical with $\bar{x}^n$.
 \begin{lemma}  Let  ${Y}^n$ be the output of a non-redundant DMC $W: {\cal X}\rightarrow {\cal Y}$ with input ${X}^n$. Then for any $x^n, \bar{x}^n\in {\cal X}^n$ with $d_H(\bar{x}^n, x^n)=\alpha n$,  any $\epsilon\in (0, {\Theta \alpha})$ and $\alpha>0$, it holds that
\begin{eqnarray}
P_{{Y}^n|{X}^n}\Big{(}\textsf{T}^n_{[W]_\epsilon}(\bar{x}^n)\Big{|} x^n\Big{)}\le  2^{-\frac{2n(\alpha\Theta-\epsilon)^2}{|{
\cal X}|^2|{\cal Y}|^2}},
\end{eqnarray}
 where  $\Theta=\min_{i}\Delta\Big{(} W_i, Cov(W_{[p]\backslash \{i\}})\Big{)}$ and the rows of $W$ are  $W_1, \cdots, W_p.$ \label{le: cap}
\end{lemma}

The following lemma   is a special case of \cite[Lemma 6] {Jiang14a}.

\begin{lemma} For  $1/n\le \alpha\le 1/2$,   there exists a subset ${V}_\alpha\subseteq  {\cal X}^n$ with $$|V_\alpha|\ge \frac{1}{\alpha n} |{\cal X}|^n 2^{-n(h(\alpha)+\alpha \log |{\cal X}|)}$$ such that $d_H(x^n_1, x_2^n)\ge \alpha n$ for any distinct $x_1^n, x_2^n\in {V}_\alpha.$ \label{le: Vd}
\end{lemma}

\subsection{$(v, b, r, \lambda)$-Set System}

We now introduce the  $(v, b, r, \lambda)$-set system in \cite{Jiang14}, which is extended from block design \cite{HP85}.
\begin{definition} Let $V$ be a set of size $v$ and ${\cal B}=\{{\cal B}_1, \cdots, {\cal B}_b\}$ (called {\em blocks}) be a collection of subsets of $V.$ Then, $(V, {\cal B})$ is a $(v, b, r, \lambda)$-{\em set system} if
\begin{itemize}
\item[1.] Each $x\in V$ belongs to at least $r$ blocks.
\item[2.] Any $x, y\in V$ simultaneously appear in at most $\lambda$ blocks.
\end{itemize} \label{def: setsys}
\end{definition}

The following  lemma is a rephrase of an  existence result proved  in \cite{Jiang14}.

 \begin{lemma} Let  $v, b, t\in \mathbb{N}$ with $b= \lfloor {\frac{2^{t+8}}{\epsilon^4}\log v}\rfloor$ and $v\ge 2$ and $0<\epsilon<1$.  Then, there exists a $(v, b, 2^{-.25t-2}\epsilon b, 2^{-.5t-2}\epsilon^2 b)$-set system. \label{le: existset}
\end{lemma}

The above lemma shows that the existence of a set system with  $b>512\epsilon^{-4}\log v$.   We now prove that $b> \log v$ actually holds for any set system with  $\epsilon<1.$ Although this result will not be directly used in this paper, it is the   main  motivation  that leads us to the  lower bound on the  round complexity in Section \ref{sect: round}.

\begin{lemma} Let $(V, {\cal B}_1, \cdots, {\cal B}_b\}$ be a $(v, b, r, \lambda)$-set system  with $\lambda<r$. Then,   $b> \log v.$
\end{lemma}
\noindent{\bf Proof. } For any $s\in V$, define a $b$-bit string $I(s)$, where the $i$th bit $I(s)_i=1$ if and only if $s\in {\cal B}_i.$ As any distinct  $s_1, s_2\in V$ simultaneously appear in at most $\lambda<r$ blocks while each of $s_1, s_2$  appears in at least $r$ blocks, it follows that $I(s_1)\ne I(s_2).$ Hence,  $I(\cdot)$ is an injection from $V$ to $\{0, 1\}^b$.
Since $I(s)$ is a $b$-bit string with at least $r$ positions being 1, $v\le  2^b-\sum_{i=0}^{r-1} {b\choose i} $. That is, $b> \log v.$ $\hfill\square$

\section{Authentication Model} \label{sect: model}

In this section, we introduce the noisy authentication model over DMCs in \cite{Jiang14}.  It consists of two DMCs:   $W_1: {\cal X}\rightarrow {\cal Z}$ from Alice to Bob and  DMC $W_2: {\cal Y}\rightarrow {\cal Z}$ from Oscar to Bob. Between Alice and Bob,  there exists  a two-way noiseless channel. Alice will use $W_1$ and the noiseless channel to authenticate a source state  to Bob. Oscar is an attacker. He can read the messages  sent over  the two-way noiseless channel and   channel $W_1$. He can also tamper the messages on the two-way noiseless channel.  Allowing Oscar to control the noiseless channel is to capture the concern that this  channel is neither  confidential nor  authenticated. Allowing   Oscar to see Alice's  message over  the DMC   $W_1$ is to capture the concern that this channel  may leak some information.  One might think that let Oscar know the full input of $W_1$ is unnecessary. However, we prefer this as it simplifies  the model and also  provides a stronger security guarantee.

After rounds  of interactions, Bob can  decide whether to accept the   authentication. When he accepts, he outputs a source state; otherwise, he outputs a special symbol $\perp$. If Bob detects an error before completing the interaction, he outputs $\perp$ and aborts  immediately.  The formal description follows.

 \subsubsection{Communication model} Let ${\cal S}$ be the source space, from which Alice draws   a  source state $S$ for  authentication. Let  $\pi_n$ be a $\nu$-round  authentication protocol with  totally $n$ symbols transmitted  over  channel $W_1$.   Each party has a basic   input and a random input (a uniformly random binary string which is the randomness source in the execution for this party). Alice's basic input is $S$ and random  input is $r_A$,  while Bob's basic input is empty and random  input is $r_B.$ If the list of messages a party has received so far is $T$, then his (or her) next action (e.g., generating a local output, an outgoing message  or making a reject/accept decision) is completely determined by his basic input, random input and $T$. We use $\pi_n(A, r_A, {T})$ to denote Alice's  next action  function and $\pi_n(B, r_B, {T})$ to denote Bob's next action function.  The interaction is as follows, where $n=\sum_{i=1}^\nu n_i.$
\begin{description}
\item[A-1: ] Alice computes $(X_1^{n_1},u_1) =\pi_n(A, S, r_A)$. She sends $X_1^{n_1}$ over channel $W_1$ and $u_1$ over the noiseless  channel, to Bob.
Oscar will see $X_1^{n_1}, Z_1^{n_1}$ and $u_1$. He can modify $u_1$ to $u_1'$. Bob will receive $Z_1^{n_1}$ from channel $W_1$ and $u_1'$ from the noiseless channel.

\item[B-1: ] Upon $Z_1^{n_1}, u_1'$, Bob computes and sends $v_1=\pi_n(B, r_B, Z_1^{n_1}, u_1')$ to Alice over the noiseless channel. Through Oscar, Alice will receive $v_1'$.

\item[]  \hspace{1.5in} \vdots

\item[A-$i$: ] Upon $v_{i-1}',$ Alice computes
\begin{eqnarray*}
(X_i^{n_i},u_{i}) =\pi_n(A, S, r_A, v_1'|v_2'|\cdots|v_{i-1}').
\end{eqnarray*}
He sends $X_i^{n_i}$ over channel $W_1$ and $u_{i}$ over the noiseless  channel.
Oscar will see $X_i^{n_i}, Z_i^{n_i}$ and $u_{i}$. He can modify $u_{i}$ to $u_{i}'$. Bob will receive $Z_i^{n_i}$ from channel $W_1$ and $u_{i}'$ from the noiseless channel.
\item[B-$i$: ] Upon $Z_i^{n_i}, u_{i}'$, Bob computes and sends
\begin{eqnarray*}
v_{i}=\pi_n(B, r_B, Z_1^{n_1}|u_1'|Z_2^{n_2}|u_2'|\cdots|Z_i^{n_i}|u_{i}')
\end{eqnarray*}
 to Alice over the noiseless channel, which, through Oscar, becomes $v_{i}'$.

\item[] \hspace{1.5in}$\vdots$

\item[B-$\nu$: ] Upon $Z_{\nu}^{n_{\nu}}, u_{\nu}'$, Bob computes
\begin{eqnarray*}
S'=\pi_n(B, r_B, Z_1^{n_1}|u_1'|Z_2^{n_2}|u_2'|\cdots|Z_{\nu}^{n_{\nu}}|u_{\nu}')
\end{eqnarray*}
for $S'\in {\cal S}\cup\{\perp\},$
where $S'=\perp$ means that he rejects the authentication while $S'\ne \perp$  means that he agrees that $S'$  is  authenticated from Alice.

\end{description}

If Alice (or Bob) detects any inconsistency  before the protocol completion, she (or he) can reject and abort the execution immediately.

\subsubsection{Security  model}  The security model is described in terms of two attacks. In a type I attack,  Oscar can change  the messages over the two-way  noiseless channel between Alice and Bob. He succeeds if Bob accepts a source state that is different from Alice's input. In a type II attack, Oscar can impersonate Alice to authenticate a source state using $W_2$ and a noiseless channel. He succeeds if Bob accepts his authentication. The formal description is as follows.

\vspace{.05in} \noindent {\em  Admissible Attacks: }
\begin{myitemize}
\item[I. ] During the  execution of $\pi_n$ between  Alice and Bob,  Oscar can see  $(X_i^{n_i}, Z_i^{n_i}, u_i)$ from  Alice and $v_i$ from Bob. He can modify $u_i$ to any $u_i'$ and $v_i$ to any $v_i'$. He succeeds if Bob outputs $S'\in  \{S, \perp\}$.

\item[II.]  Oscar can impersonate Alice to execute $\pi_n$ with Bob, except that the noisy channel $W_1$ is replaced by $W_2.$ He succeeds in this attack  if Bob outputs $S'\ne \perp.$

\end{myitemize}
We use {\em succ} to denote a success event in a type I or II attack.

\vspace{.05in} \noindent {\em Security definition: }\quad In this paper, we assume by default  that an honest Alice (or Bob) follows the protocol with a random input that is a uniformly random binary string. However, we also consider an honest Alice or Bob who follows the protocol specification with some $r\in \{0, 1\}^*$ as the random input. In this case, we call her (or him) an {\em admissible} user.
Now the security consists of two properties: correctness and authentication. The correctness requires that  if an admissible Alice authenticates $S$ to Bob when  no attack is performed, Bob should output $S'=S$. The authentication requires that Oscar will never succeed in a  type I or II attack.

\begin{definition} An  authentication protocol $\pi_n$ for source ${\cal S}$ is  {\em secure} if it satisfies two properties.
\begin{itemize}
\item[\textbullet] \textsf{Correctness. }  For any admissible Alice, Bob outputs $S'\ne S$ only negligibly (in $n$) if  no attack is performed.
\item[\textbullet] \textsf{Authentication. } Under type I and type II attacks,  $\Pr(succ)$ is negligible in $n$.
\end{itemize}
\end{definition}

Note that here we require the error probability to be {\em negligible} (see  Section \ref{sect: symbol}) as  this   is the widely accepted quantity  for a probabilistic  event that is unlikely to occur.
\subsubsection{Authentication rate and authentication capacity}

We regard the noisy channel as an expensive resource and   the noiseless channel as a cheap source.  So we are interested in maximizing the efficiency of channel $W_1$ and define  the {\em authentication rate} of $\pi_n$ as  the ratio $\frac{\log |{\cal S}|}{n}$. The authentication model with   $(W_1, W_2)$ has an  {\em authentication capacity} $C_{a}$, if any authenticate rate $r<C_{a}$ can be achieved by a certain protocol $\pi_n$ while any protocol with an authentication rate $r>C_a$ is insecure.

\section{Our authentication protocol}
This section extends the 3-round authentication protocol in \cite{Jiang14}. The number $n$ of channel $W_1$ uses in  the  protocol of \cite{Jiang14} satisfies $n\ge \frac{\log\log|{\cal S}|}{C}$, where $C$ is the shannon capacity of $W_1$ and  ${\cal S}$ is the source space. In this section, we improve the protocol such that $n$ does not depend on  $|{\cal S}|$ but with the price that the   round complexity is  $\log^*|{\cal S}|-\log^*n+4$. Under our result, the authentication rate $\frac{\log|{\cal S}|}{n}$ is proportional to $\log|{\cal S}|.$

The  3-round protocol in \cite{Jiang14} is based on  a set system $({\cal S}, {\cal B}_1, \cdots, {\cal B}_b)$. The idea is as follows.  Alice first sends the source state $S$ to Bob noiselessly. Bob then finds all possible $i$'s such that $S\in {\cal B}_i$ and picks a random ${\cal B}_j$ among them and sends $j$ to Alice noiselessly. Finally, Alice sends $j$ via DMC $W_1$ to Bob. The construction is designed such that if $S$ is modified to $S'$ by Oscar, then a successful type I  attack implies  $S', S\in {\cal B}_j$, which is unlikely due to the property of the set system.

Our new protocol stems from \cite{Jiang14} with the following idea. Essentially, Alice still attempts to authenticate $S$ using a set system $({\cal S}, {\cal B}_1, \cdots, {\cal B}_b)$. However, she does not sends $j$ over $W_1$. Instead, he regards $j$ as a new source state in a new but  smaller source space ${\cal S}'=[b]$ and attempts to use a smaller set system $({\cal S}', {\cal B}_1, \cdots, {\cal B}_{b'})$ to authenticate $j$. It is important to notice that $b$ has the order of $\log |{\cal S}|$ by Lemma \ref{le: existset}. Similarly,  $b'$ has the order of $\log b$, which in turn has   the order of $\log\log |{\cal S}|.$ So to authenticate $j$, Alice now only needs to send a DMC message from a domain of $b'=\log\log|{\cal S}|$ (instead of a domain of size $b=\log|{\cal S}|$). That is, two iterations on the protocol of \cite{Jiang14} allow to decrease DMC message to the log size. Continuing with this idea, if we iterate the protocol \cite{Jiang14} for $L$ times, then conceivably  the DMC message will reduce to a domain size of $\log^{(L)} |{\cal S}|.$ Thus, if the DMC message length  is  $n$, then it suffices  to iterate the protocol in \cite{Jiang14} for  $\log^*|{\cal S}|-\log^*n+O(1)$ times (using the fact $\log^* m=L+\log^*(\log^{(L)} m)$ for any $m$ and $L\le \log^*m$). This gives our desired result.   In the  following, we implement this  idea    rigorously.

\subsection{The construction}
For any $R<C$ ($C$ is the shannon capacity over $W_1$), Shannon capacity theorem tells us that there exists a channel code ${\cal C}=\{C_1, \cdots, C_{2^{n'R}}\}\subseteq {\cal X}^{n'}$ that has  a  maximum error probability $\delta_{n'}\rightarrow 0$ {\em exponentially} with $n'$ (see \cite{CK11}). Assume  ${\cal C}$ has an encoding $f_{n'}$ and a decoding $g_{n'}$.

For $v_1\in \mathbb{N}$, let ${\cal S}=[v_1]$ be the source space.  Take $\epsilon=2^{-\beta_1 n'}$ for some $\beta_1\in (0, R/4).$
 Let  $\phi$ be the minimal {\em even} $t$ s.t.  $\log^{(t)}v_1\le \beta_2 n'+\sqrt{n'}$ for some $\beta_2\in (0, R-4\beta_1)$.  Let $v_{j+1}=\lfloor\frac{2^{\phi-j+8}}{\epsilon^4}\log v_{j}\rfloor$ for each $j<\phi$. By Lemma \ref{le: existset}, there exists a $(v_{j}, v_{j+1}, 2^{-.25(\phi-j)-2}\epsilon v_{j+1}, 2^{-.5(\phi-j)-2}\epsilon^2v_{j+1})$-set system, which we denote by  $\mathbb{S}_{j}=({\cal S}_{j}, {\cal B}_{j,1}, \cdots, {\cal B}_{j, v_{j+1}})$ with  ${\cal S}_{j}=[v_{j}]$. Assume Alice wants to authenticate source state $s\in {\cal S}. $ The protocol is described in Fig. \ref{fig: SetAuth}, where we assume $W_1(\cdot|a)\not\in Cov(W_2)$ for some $a\in{\cal X}$.
\begin{figure}[ht!]
\frametext{
\begin{itemize}
\item[0.] Let $s_1=s, s_0=1, L_{0,i}=\{1\}$ for any $i\in [v_1].$
\item[1.]  For $\ell=1$ to $\phi$, do the following. Let ${\cal P}_1={\cal P}_3=\cdots=$Alice and ${\cal P}_2={\cal P}_4=\cdots=$Bob.
\begin{itemize}
\item[a.]  ${\cal P}_\ell$  sends $s_\ell$  to ${\cal P}_{\ell+1}$ over the noiseless channel, which, through Oscar,  arrives at ${\cal P}_{\ell+1}$ as $s'_{\ell}$.
\item[b.] Upon $s'_{\ell}$,  ${\cal P}_{\ell+1}$ checks if $s_{\ell-1}\in {\cal B}_{\ell-1, s_{\ell}'}$. If not, (s)he rejects; otherwise,  (s)he determines   $$L_{\ell}=\{{\cal B}_{\ell, i}\mid s'_\ell\in {\cal B}_{\ell,i}, i\in [v_{\ell+1}]\}.$$ Assume $L_{\ell}=\{{\cal B}_{\ell, i_1}, \cdots, {\cal B}_{\ell, i_r}\}$  ($r$ might vary with $s'_\ell$). If $\ell<\phi$, (s)he takes $s_{\ell+1}$ from  $\{i_1, i_2, \cdots, i_r\}$ uniformly randomly and proceeds to iteration  $\ell+1$; otherwise ($\ell=\phi$, even, $P_{\ell+1}=$Alice), she goes to step 2.
\end{itemize}
\item[2.]  Alice  sends $C_{s_{\phi}'}^*=a^k|C_{s_{\phi}'}$ over $W_1$ for $k=\sqrt{n'}$.
\item[3.] Upon  $Z^{n'+k}$, Bob checks  if $$|T_{Z^k}(u)-W_1(u|a)|\le \frac{\gamma}{2|{\cal Z}|}$$ for all $u\in {\cal Z}$, where $\gamma=\Delta(W_1(\cdot|a); Cov(W_2))$.  If no, he rejects; otherwise, he accepts if and only if $Z_{k+1}^{k+n'}$ is decoded to $s_{\phi}$.
\end{itemize}
}
\caption{Our authentication protocol   {\tt SetAuth}$^*$} \label{fig: SetAuth}
\end{figure}
\subsection{Security analysis}
Now we analyze the security of our new protocol. Before this, we first prove the following preparation lemma.

\begin{lemma} Let $0<\delta<1, k\in \mathbb{N}, v_1>0$ with  $\log^{(k)}v_1\ge 3.$  If $v_{j+1}\le \frac{2^{k-j}}{\delta}\log v_{j}$ for $1\le j\le  k,$  then $$v_{j+1}<\frac{2^{k-j}\log^{(j)} v_1}{\delta}+\frac{2^{k-(j-1)}}{\delta}\log(\frac{2^{k-(j-1)}}{\delta}).$$ \label{le: round}
\end{lemma}
\noindent{\bf Proof. }
The conclusion holds for the initial case $j=0$ automatically. Assume it holds for case $j-1$. Consider case $j$. Let $\alpha_i=\frac{2^{k-(i-1)}}{\delta}\log(\frac{2^{k-(i-1)}}{\delta})$ for any $i$.  By induction,

\vspace{.05in}  \noindent $v_{j+1}<\frac{2^{k-j}}{\delta}\log(\frac{2^{k-(j-1)}\log^{(j-1)} v_1}{\delta}+\alpha_{j-1})$

\vspace{.05in}\hspace{.12in}  $\le  \frac{2^{k-j}}{\delta}\log(\frac{2^{k-(j-1)}\log^{(j-1)} v_1}{\delta})+\frac{\alpha_{j-1}}{2\ln2\cdot \log^{(j-1)} v_1}$

\vspace{.05in} \hspace{.12in} $\le   \frac{2^{k-j}}{\delta}\log^{(j)} v_1+\frac{2^{k-j}}{\delta}\log(\frac{2^{k-(j-1)}}{\delta})+\frac{\alpha_{j-1}}{\log^{(j-1)} v_1}$

\vspace{.05in} \hspace{.12in} $\stackrel{(*)}{\le}  \frac{2^{k-j}}{\delta}\log^{(j)} v_1+2*\frac{2^{k-j}}{\delta}\log(\frac{2^{k-(j-1)}}{\delta})$

\vspace{.05in}  \hspace{.12in} $=   \frac{2^{k-j}}{\delta}\log^{(j)} v_1+\alpha_{j},$

\vspace{.05in} \noindent  where inequality $(*)$ uses the fact that $\log^{(j-1)} v_1\ge \log^{(k-1)}v_1\ge 2^3$ and that $4x\log x\ge 2x\log(2x)$ for $x=\frac{2^{k-(j-1)}}{\delta}\ge 2$.
$\hfill\square$

Applying the lemma to our construction with $\delta=2^{-8}\epsilon^4$ and $k=\phi-1$, we have
\begin{corollary} Keep notions as in protocol {\tt SetAuth}$^*$. Then,
\begin{equation}
v_{\phi}<2^{9+4\beta n'}\Big{(}\log^{(\phi-1)} v_1+{20+8\beta n'}\Big{)}.
\end{equation} \label{cor: vphi-1}
\end{corollary}

 \begin{theorem}  If $Cov(W_1)\not\subseteq Cov(W_2)$ and $\dim W_1>1$,  then {\tt SetAuth}$^*$ is a $2^{-\xi \sqrt{n}}$-secure authentication protocol for a constant $\xi>0$ with round complexity at most $\log^* v_1-\log^*n+4$, where $n=n'+\sqrt{n'}$ is  the number of channel $W_1$ uses which  does not  depend  on $v_1$.  \label{thm: SetAuth}
\end{theorem}

\noindent {\bf Proof. } {\em Correctness. } When Oscar does not involve in the attack, $s_\ell=s_\ell'$ for all $\ell.$ From Corollary \ref{cor: vphi-1} and $4\beta_1+\beta_2<R$, we know that $v_{\phi}<2^{n'R}$ when  $n'$ large enough.   Since $s_{\phi}$ is taken from ${\cal S}_{\phi}$ (of size $v_{\phi}$),  Bob will decode $C_{s'_\phi}$ to $s'_\phi$ with an exponentially small error probability, by the assumption of ${\cal C}$. In addition, by Lemma \ref{le: basicH}, $|T_{Z^k}(u)-W_1(u|a)|\le \frac{\gamma}{2|{\cal Z}|}$ for all $u\in {\cal Z}$
 is violated with an exponentially small probability too.  The correctness follows.

\noindent{\em Authentication. }
\quad  By the authentication model, there are two types of attacks.
\begin{myitemize}
\item[{\bf Type-I.} ] Oscar revises messages  over the noiseless channel between Alice and Bob such that $s_1 \ne s_1'$.
\item[{\bf Type-II.} ] Oscar plays the role of Alice to interact with bob to authenticate $\tilde{s}$, where assume that the message in the iteration $\ell$ in step 1 is $\tilde{s}_\ell.$ Further, at step 2,  we assume Oscar sends $\tilde{C}^*$ over the channel $W_2$  to Bob.
\end{myitemize}

For a type I attack, the success probability  is bounded by $P({succ}|s_\phi'\ne s_\phi)+P({succ}|s_\phi'=s_\phi).$ Note that $succ$ event implies the decoding result $g_{n'}(Z_{k+1}^{k+n'})=s_\phi$. By correctness of code ${\cal C}$, $P(g_{n'}(Z_{k+1}^{k+n'})=s_\phi')>1-2^{-n'\alpha}$ for some $\alpha>0$. So $P({succ}|s_\phi'\ne s_\phi)\le 2^{-n'\alpha}.$ We thus focus on the case $s_\phi'=s_\phi.$ In this case, as $s_1'\ne s_1$, there  must exist $j<\phi$ such that $s_j'\ne s_j$ but $s_{j+1}'=s_{j+1}$. In this case, notice that $P_{j+2}$ will verify whether $s_j\in {\cal B}_{j, s_{j+1}'}$. We now bound the probability for this to hold. First, observe that the time order for  $s_j, s_j', s_{j+1}=s_{j+1}'$ is as follows:
$P_j$ generates $s_j$; then, Oscar revises it to $s_j'$; next, upon $s_j'$, $P_{j+1}$ generates $s_{j+1}$; finally,  $P_{j+2}(=P_j)$ receives $s_{j+1}'=s_{j+1}.$  Thus, $s_{j+1}=s_{j+1}'$ is selected after $s_j$ and $s_j'$ have been fixed.  By the definition of  $s_{j+1}$, it holds that $s_{j}'\in {\cal B}_{j, s_{j+1}}$. Since $P_{j+2}$ will verify  $s_j\in {\cal B}_{j, s_{j+1}'}$, it follows that a successful attack implies $s_j, s_j'\in {\cal B}_{j, s_{j+1}}.$
However, as  $s_{j+1}$ is uniformly randomly from $\{i_1, \cdots, i_r\}$, this probability is at most $2^{-.25(\phi-j)}\epsilon$, by the property of the set system $\mathbb{S}_{j}$. Since $j$ can take any value from 1 to $\phi-1$, it follows that
 $$P({succ}|s_\phi'=s_\phi)\le \sum_{j=1}^{\phi-1} 2^{-.25(\phi-j)}\epsilon< \frac{2^{-.25}\epsilon}{1-2^{-.25}}<6\epsilon.$$   Hence, a type I attack succeeds with probability at most $2^{-n'\alpha}+6\epsilon,$ which is exponentially small as $\epsilon=2^{-\beta n'}.$

  For a type II attack, assume  Bob receives $Z^{n'+k}$.   We claim    $$P_{Z^k}\Big{(}|T_{Z^k}(u)-W_1(u|a)|\le \frac{\gamma}{2|{\cal Z}|}, \forall u\Big{)}\le 2e^{-\frac{k\gamma^2}{8|{\cal Z}|^2}}.$$ Otherwise, by Lemma \ref{le: neg},
$$\Delta\Big{(}W_1(\cdot|a); Cov(W_2)\Big{)}\le \gamma/2+\gamma/4<\gamma. $$ This is impossible, as  $\Delta(W_1(\cdot|a); Cov(W_2))=\gamma$.
This completes the proof of the authentication property by defining $\xi<\frac{\gamma^2}{8|{\cal Z}|^2}.$

Finally, as $\log^*v_1=\phi+\log^*(\log^{(\phi)}v_1)$ and $\log\log (\beta_2n)<\log^{(\phi)}v_1\le n$ by the definition of $\phi$, we have $\phi\le \log^*v_1-\log^* n+3$ (using $2^{\beta_2 n}\ge n$) for $n$ large enough.  This gives the  round complexity.  $\hfill\blacksquare$
\section{Lower Bound on Round Complexity} \label{sect: round}

In this section, we prove a lower bound on the round complexity of an authentication protocol in our model. Our strategy is to reduce the problem to a special class of protocols and then bound the round complexity of the latter.

Toward this, we  define $\Sigma_1$ to be the set of authentication protocols in our model such that  the DMC message over $W_1$ is sent only in the final flow and the final flow has no message over the noiseless channel.

In the following, we  show that if there is  an $L$-round secure authentication protocol in our model, there exists a secure $L'$-round protocol in $\Sigma_1$ with  $L'\le L+2$. Our idea is that we can move  each DMC message $X^{n_i}$ in the original protocol to the noiseless channel of the same flow  and in addition  also   send $X^{n_i}$ over DMC $W_1$ in the final flow.  This modification needs to be careful:  the original protocol could use the DMC output $Y^{n_i}$ right after Bob has  received it  while the modified  protocol only has the noiseless version $X^{n_i}$ (instead of  $Y^{n_i}$). Fortunately, this can be fixed by permitting   Bob to simulate      $Y^{n_i}$ (letting ${X'}^{n_i}$ go through a statistical   model that has the same characteristics as channel $W_1$), where ${X'}^{n_i}$ is the received version of $X^{n_i}$ by Bob over the noiseless channel. However, this causes a new problem:  it is possible that ${X'}^{n_i}\ne X^{n_i}$. To overcome  this, we actually send $X^{n_i}$ in the final flow  using an error-correcting code, through  which Bob can obtain  $X^{n_i}$ with high probability. In addition, $X^{n_i}$ is coded such that if ${X'}^{n_i}\ne X^{n_i}$, then  the change  can be  detected. The formal result is as  follows.

\begin{lemma}
If there exists an $L$-round  $\epsilon$-secure authentication protocol $\pi$  in our  model, then there exists an $L'$-round $(\epsilon+2^{-\beta n'})$-secure authentication protocol $\pi'\in \Sigma_1$ with  $n'={\mu n}$ for $L'\le L+2$ and some constants  $\beta>0, \mu>0$, where  $n', n$ are respectively  the numbers of channel $W_1$ uses in $\pi', \pi$. \label{le: finalW1}
\end{lemma}

\noindent{\bf Proof. } Let $\pi$ be an $L$-round $\epsilon$-secure authentication protocol in our model. We construct an $L'$-round $(\epsilon+2^{-\beta n'})$-secure authentication protocol $\pi'$ from $\pi$ as follows. W.L.O.G., assume $W_1(\cdot|a)\not\in Cov(W_2)$ (by \cite{Jiang14}, a necessary condition for $\epsilon$-secure authentication is $Cov(W_1)\not\subseteq Cov(W_2)$).
\begin{itemize}
\item[i.] Alice follows $\pi$, except that whenever she needs to send $F$ over $W_1$, she instead sends it over the noiseless channel.
 \item[ii.] Bob follows $\pi$, except that whenever he receives $F'$ over the noiseless channel (the received version of $F$, where $F$  is  supposedly sent over DMC $W_1$ in $\pi$), she lets  it go through a simulated  $W_1$ and regards  the output  as the DMC output  in $\pi$ and  proceeds  normally according to $\pi$.
 \item[iii.] If the $L$th flow in $\pi$ is from Alice to Bob, then Bob sends $0$ as the $(L+1)$th flow in $\pi'$ and the $(L+2)$th flow will be the final flow; otherwise, the $(L+1)$th flow will be the final flow. In any case, the final flow in $\pi'$ is from Alice to Bob and defined as follows.   Let $(F_1, \cdots, F_L)$ be the list of messages that are sent over DMC $W_1$ in $\pi.$ Since $\pi$ uses $W_1$ for $n$ times, it follows $F^L\in {\cal X}^n$. Let $\bar{n}=\frac{2n\log|{\cal X}|}{C}$, where  $C$ is the shannon capacity of $W_1$ ($C>0$ is implied by the necessary condition $\dim W_1>1$  \cite{Jiang14}). By Shannon capacity theorem, there exists a code ${\cal C}\subseteq {\cal X}^{\bar{n}}$ over channel $W_1$ for source ${\cal M}={\cal X}^n$ that has an exponentially small error probability (say, $2^{-\alpha \bar{n}}$ for some $\alpha>0$).  Alice encodes $(F_1, \cdots, F_L)$ to $X^{\bar{n}}\in {\cal C}$ and sends $a^{\bar{n}}X^{\bar{n}}$  over DMC $W_1$ in the final flow of $\pi'$.
 \item[iv.]  Let $Y^{2\bar{n}}$ be the received vector in the final flow $\pi'$ for $a^{\bar{n}}X^{\bar{n}}$ over channel $W_1$. Bob will accept the authentication if and only if

     \begin{itemize}
     \item the original verifications in $\pi$  are satisfied;
     \item $Y_{\bar{n}+1}^{2\bar{n}}$ decodes  to ${F'}^{L}$ (the received version of $F^L$ over the noiseless channel by Bob in $\pi'$);
      \item $Y^{\bar{n}}\in \textsf{T}_{[W_1]_{.5\gamma}}^{\bar{n}}(a^{\bar{n}})$ for  $\gamma=\Delta(W_1(\cdot|a), Cov(W_2))$.
      \end{itemize}
\end{itemize}
This completes the description of $\pi'$.

Now we analyze the security of  $\pi'.$
Consider a type I attack first. For any  Oscar$'$ against $\pi'$ (executed between Alice$'$ and Bob$'$), we construct Oscar against $\pi$ (executed between Alice and Bob). The strategy of Oscar is to maintain a simulated Alice$'$ and Bob$'$ to execute $\pi'$ with Oscar$'$ against it and then mimic the attack strategy  of Oscar$'$ to attack $\pi$. Toward this, the simulation of  Alice$'$ and Bob$'$ will rely on the view of Oscar in the execution of $\pi.$ Details follow.
\begin{itemize}
\item[-] When Alice (or Bob) in $\pi$ sends $M$ over the noiseless channel, Oscar lets Alice$'$ (or Bob$'$) does the same thing in $\pi'$ and also lets Oscar$'$ know $M$. In addition, whenever  Alice sends $F_i$ over channel $W_1$, Oscar lets Alice$'$ in $\pi'$ sends $F_i$ to Bob$'$ over the  noiseless channel.
 \item[-] When Oscar$'$ (against $\pi'$) changes $M$ to $M'$ before the delivery, Oscar (against $\pi$) does the same thing. When Oscar$'$ changes $F_i$ to $F_i'$, Oscar aborts immediately; otherwise, Oscar$'$ will deliver $F_i$ without a change (recall that Alice in $\pi$ has sent $F_i$ over $W_1$). If   Bob in $\pi$  receives $\bar{Y}^{n_i}$ over $W_1$ (when Alice sends $F_i$), then Oscar lets Bob$'$ use $\bar{Y}^{n_i}$ as the simulated output of $W_1$ with input $F_i$. Note this $\bar{Y}^{n_i}$ has the same distribution as the simulated output by  Bob$'$ in $\pi'$ as they are both according to the statistic model $W_1$.
 \item[-] In the last round of $\pi'$,  Oscar simulates Alice$'$ and Bob$'$ to act normally. He  lets Oscar$'$ know the input $a^{\bar{n}}X^{\bar{n}}$  and     output $Y^{2\bar{n}}$ of DMC   $W_1$.
\end{itemize}
Denote the above attack of Oscar by $\Gamma'$. Note that the view of Oscar$'$ in $\Gamma'$ is according to the distribution in a real attack. It suffices to bound the success event (denoted by succ$'$) of Oscar$'$ in $\Gamma'$. Thus, $$P(succ')=P(succ', F_i\ne F_i', \exists i)+P(succ', F^L= {F'}^L).$$
 Note if $(F_1, \cdots, F_L)\ne (F_1', \cdots, F_L')$, $succ'$ implies a decoding error for $Y_{\bar{n}+1}^{2\bar{n}}$, which is bounded by $2^{-\alpha \bar{n}}$ for some $\alpha>0$ (by the classic random coding result as the information rate is less than $\frac{\log |{\cal X}|^n}{\bar{n}}\le C/2<C$). Further, when $(F_1, \cdots, F_L)= (F_1', \cdots, F_L')$, the success of Oscar$'$ in $\pi'$ implies the success of Oscar in $\pi$, which is bounded by $\epsilon$ due to our assumption for $\pi$. Hence, $P(succ')\le 2^{-\alpha \bar{n}}+\epsilon.$

Then, we consider type II attack. In this case, it is similar to the analysis of type II attack in SetAuth$^*$ that the success probability of the attacker is bounded by $2e^{-\frac{\bar{n}\gamma^2}{8|{\cal Z}|^2}}.$

As a summary, the success probability of type I and II attacks is  bounded by $\epsilon'=\epsilon+2^{-\alpha \bar{n}}+2e^{-\frac{\bar{n}\gamma^2}{8|{\cal Z}|^2}}.$ Finally, the number of channel $W_1$ uses in $\pi'$ is  $n'=2\bar{n}=\frac{4n\log|{\cal X}|}{C}$. Thus,  a value is negligible in $n'$ if and only if it is negligible in $n$. Thus, $\pi'$ is $\epsilon'$-secure under parameter $n'$.  This completes our proof. $\hfill\blacksquare$

\vspace{.10in} In the following, we show that we can always assume the first flow of the protocol is  the source state $S$ over the noiseless channel from Alice. The idea is that the source state is not confidential and hence the authentication property does not depend  on its secrecy. Thus, if it is not sent in the first flow, then we can prepend it  to  the protocol.
\begin{lemma} Let $\pi$ be an $L$-round  $\epsilon$-secure authentication protocol in our model for source space ${\cal S}$. Let   $\pi'$ be  an authentication protocol obtained from $\pi$ as follows:
\begin{itemize}
\item The first flow of $\pi'$ is  the source state $S$ over the noiseless channel from Alice;
\item If the first flow in $\pi$ is from
Alice, then the second flow of $\pi'$ is a constant message 0 over the noiseless channel from Bob;
\item After the preliminary   flow(s) above, Alice and Bob start to execute $\pi$ normally with $S$ as  Alice's input in $\pi$.
 \end{itemize}
 Then, $\pi'$ is an $L'$-round $\epsilon$-secure authentication in our model with  $L'\le L+2$.  \label{le: startM}
\end{lemma}
\noindent{\bf Proof. }  If there exists an Oscar$'$ against $\pi'$, we present an Oscar against $\pi.$ We describe Oscar for type I and II attacks as follows.  Assume $\pi$ is run between Alice and Bob and $\pi'$ is run between Alice$'$ and Bob$'.$ The strategy of Oscar is to  simulate Alice$'$ and Bob$'$ and run Oscar$'$ against the execution of $\pi'$. W.L.O.G., assume $\pi$ starts with Alice.

For a type  I attack,  Oscar does as follows.
\begin{itemize}
\item When  Oscar$'$ invokes Alice$'$  (in $\pi'$) to authenticate $S$ to Bob$'$, Oscar simulates Alice$'$ with input $S$ and sends $S$ to Bob$'$, which through Oscar$'$ will be delivered to Bob$'$ as $S'.$ Bob$'$ will then send $0$ to Alice$',$ which we assume to  arrive at Alice$'$ as 0 (otherwise, Alice$'$ simply rejects).  In this case, Oscar invokes Alice (in $\pi$) with input $S$. Further,  Oscar simulates Alice$'$ and Bob$'$ to start $\pi$ (as a subprotocol of $\pi'$) with input $S$,  by strictly following the flows  between Alice and Bob. Details follow.
\item  Whenever Alice (or Bob) sends a message $C$ to Bob  (or Alice) noiselessly, Oscar simulates Alice$'$ (or Bob$'$) to send $C$ to Bob$'$ (or Alice$'$)  noiselessly as well.
 \item Whenever Oscar$'$ delivers a message $M'$ to Bob$'$ (or Alice$'$), Oscar delivers $M'$ to Bob (or Alice)  in $\pi$ as well.
 \item Whenever Alice sends a message $X^t$ to Bob over $W_1$, Oscar simulates Alice$'$ to send  $X^t$ over (virtual) $W_1$  as well and informs Oscar$'$ about this. When $X^t$ in $\pi$ arrives at Bob as $Y^t$, Oscar delivers $Y^t$ to Bob$'$ as the output of $W_1$ and also notifies  $Y^t$ to Oscar$'$.
\end{itemize}
From the description of Oscar, the view of Oscar$'$ is distributed according to the real attack. Also when Oscar$'$ successfully authenticates $S'\ne S$ to Bob$'$, Oscar do so to Bob as well,  as the execution of $\pi$ between Alice$'$ and Bob$'$ and the execution of $\pi$ between Alice and Bob are identical. Especially, Bob$'$ accepts $S'$ if and only if Bob accepts $S'.$ Thus, Oscar has  the same success probability as  Oscar$'$.

For type II attack, Oscar's strategy is similar, omitted. $\hfill\blacksquare$

\vspace{.10in} In the following, we will prove our lower bound on the round complexity.  Our  idea is as follows.
By Lemma \ref{le: finalW1} and Lemma \ref{le: startM}, we only need
to consider a protocol $\pi$ whose first flow is the source state $S$ over the noiseless channel from Alice and  the final flow consists of only a DMC message from Alice, which also is  the only flow that has  a DMC message. We first consider such a protocol of  3-round and show that its source space must be bounded by $2^{1+2^{|{\cal X}|^n+1}}$. If $u^{j-1}$ is  the first $j-1$ flows, then we  define  ${\cal M}_j(u^{j-1})$ to be  the set of all possible messages in the $j$th flow. For convenience, we regard  \textsf{reject} is  also as a possible message.  It is immediate that ${\cal M}_3(u^2)\subseteq {\cal X}^n\cup \{\textsf{reject}\}.$ If we sort  ${\cal X}^n\cup \{\textsf{reject}\}$ in any fixed order, ${\cal M}_3(u^2)$ can be represented by a binary vector $\textsf{D}(u^2)=(d_0, \cdots, d_{|{\cal X}|^n})$, where $d_i=1$ if and only if ${\cal M}_3(u^2)$ contains the $i$th element in  ${\cal X}^n\cup \{\textsf{reject}\}$. Thus, each $\textsf{D}(u^2)$ must be one of these $2^{|{\cal X}|^n+1}$ binary vectors.  Now we consider the case where the second flow $u_2$ is always  0 (constant). In this case, if $|{\cal S}|>2^{|{\cal X}|^n+1}$, then there must exist $u_1, \bar{u}_1$ such that $\textsf{D}(u_10)=\textsf{D}(\bar{u}_10).$ Then, Oscar can attack $\pi$ as follows. He first requests Alice to authenticate $u_1$ and then  modifies the first flow  $u_1$ to $\bar{u}_1$ but keeps  other flows unchanged. Under this attack, Oscar is admissible,  as ${u}_3\in {\cal M}_3(\bar{u}^2)$ from  $\textsf{D}(u_10)=\textsf{D}(\bar{u}_10)$). By the correctness of the authentication protocol, Bob will accept $\bar{u}_1$ and hence Oscar succeeds. This contradicts the authentication property. Thus, we must have  that  $|{\cal S}|\le 2^{|{\cal X}|^n+1}$. Our foregoing argument is based on the restriction that $u_2$ is a constant, which is of course not true usually.  However, for the general case, we might  still wish to use a certain variant of this strategy. Specifically, we may try  to define $\textsf{D}(u)$ such that if the number of possible vectors $\textsf{D}(u)$ is  less than $|{\cal S}|$, then there must exist  two source states $u_1, \bar{u}_1$ which share the same possible  choices for the second flow and the third flow. In this case, the above attack can go through. Toward this, we use $\mathbb{D}_2$ to denote all possible $\textsf{D}(u^2)$ and define $\textsf{D}(u_1)=(d_0, \cdots, d_{|\mathbb{D}_1|})$, where $d_i=1$ if and only if there exists $u_2$ such that $\textsf{D}(u^2)$ is the $i$th element in $\mathbb{D}_2\cup \{\textsf{reject}\}$. Notice that $|\mathbb{D}_2|\le 2^{|{\cal X}|^n+1}$. Hence, under our  treatment, an variant of Oscar's attack above succeeds  if the number of all possible $\textsf{D}(u)$ is less than $|{\cal S}|$ (which is guaranteed if $|{\cal S}|>2^{1+2^{|{\cal X}|^n+1}}$, or roughly $\log^{(2)}|{\cal S}|>|{\cal X}|^n$).  So the authentication property necessarily implies $\log^{(2)}|{\cal S}|\le |{\cal X}|^n$ (roughly). For a general $L$-round protocol, we can generalize the above   idea to show that  $\log^{(L-1)}|{\cal S}|\le |{\cal X}|^n$ (roughly).
From $L-1=\log^*{|{\cal S}|}-\log^*(\log^{(L-1)} |{\cal S}|),$ this gives  $L-1\ge \log^*{|{\cal S}|}-\log^*(|{\cal X}|^n)$, which is basically our desired lower bound on the round complexity.  We now implement the above idea rigorously. We start with a claim.

\vspace{.10in} \noindent {\bf Claim 1. } \quad  If $D_L\ge 3$ and $D_i\le  2^{1+D_{i+1}}$ for any $i=1, \cdots, L-1$, then $\log^{(L)} D_1\le 1+\log D_L.$

\vspace{.05in} \noindent {\bf Proof. } It suffices to prove the bound when $D_i= 2^{1+D_{i+1}}$ for each $j$, as in this case $D_1$ achieves the largest possible value.  Notice that  if $\log A_1\le b+A_2$ for $A_2\ge 3$, then
 \begin{align}
\log^{(2)} A_1\le \log A_2+\log (1+\frac{b}{A_2})\le \log A_2+\frac{b}{2}.\label{eq: Claim-1}
 \end{align}
Hence, from $\log D_j= 1+D_{j+1}$ and $D_{j+1}\ge 3$ (as $D_L\ge 3$), 
  \begin{align*}
  \log^{(2)} D_1\le \log D_2+\frac{1}{2}\le  D_3+1+\frac{1}{2}
\end{align*}
Using Eq. (\ref{eq: Claim-1}) again, we have
  \begin{align*}
  \log^{(3)} D_1\le \log D_3+(1+\frac{1}{2})/2\le D_4+1+\frac{1}{2}+\frac{1}{2^2}.
\end{align*}
Continuing this evaluation, we have
  \begin{align*}
  \log^{(L)} D_1\le& \log D_L+(1+\frac{1}{2}+\cdots+\frac{1}{2^{L-1}})/2\\
  \le& \log D_L+1.  
\end{align*}
This completes the proof.  $\hfill\blacksquare $

  We now formally present our theorem.
\begin{theorem} Let $\pi$ be an $L$-round  $\epsilon$-secure authentication protocol for source space ${\cal S}$. Then $L\ge \log^*|{\cal S}|-\log^*{n}-5$,  where $n$ is the number of channel $W_1$ uses.   \label{thm: RC}
\end{theorem}
\noindent{\bf Proof. } We first  prove the theorem for  $\pi$ with the following restrictions: (a)  the first flow is the source state $S$ over the noiseless channel from Alice; (b)   the final  flow is a DMC message over $W_1$  from Alice  and  a DMC message is only sent in the final flow.

If the first $j-1$ flows are $u^{j-1}$, we define ${\cal M}_{j}(u^{j-1})$ to be the set of possible messages in the $j$th flow by $U\in\{$Alice, Bob$\}$. Formally, $u_j\in {\cal M}_{j}(u^{j-1})$ if and only if there exists random tape $r$ such that  the list of messages of $U$ with random tap $r$ (given the list of incoming message $u_{j-1}, u_{j-3}, \cdots$) are $u_j, u_{j-2}, \cdots.$ For convenience, if  $U$ rejects (given $u^{j-1}$), we regard it as  $u_j=\perp$, where  $\perp$ is  different from any legal message flow. When $U$ rejects, (s)he aborts the execution immediately. Since $\perp$ is not an actual message flow, $u_j=\perp$ will be never delivered. Hence, when $U$ has the view of $u^{j-1}$ on the first $j-1$ flows and is going to compute $u_j$, then implicitly $u_i\ne \perp$ for any $i\le j-1.$

 By the definition of $n$, we have $u_L\in {\cal X}^n\cup\{\perp\}.$ If $u^{L-1}$ is the first $L-1$ flows, then we define a $({|{\cal X}|^n}+1)$-dimensional binary vector $\textsf{D}(u^{L-1})=(d_0, d_1, \cdots, d_{|{\cal X}|^n})$, where $d_t=1$ if and only if the $t$th element in ${\cal X}^n\cup\{\perp\}$ (sorted in any fixed order) belongs to  ${\cal M}_L(u^{L-1})$. Define $\mathbb{D}_{L-1}=\{\textsf{D}(u^{L-1})\mid \mbox{$u^{L-1}$ over all possible choices for  the first $L-1$ flows}\}.$ It is immediate that $|\mathbb{D}_{L-1}|\le 2^{|{\cal X}|^n+1}$. Now if $\textsf{D}(u^j)$ and $\mathbb{D}_{j}$ is well-defined, we define $\textsf{D}(u^{j-1})$ and $\mathbb{D}_{j-1}$. Define $\textsf{D}(u^{j-1})=(d_0, d_1, \cdots, d_{|\mathbb{D}_j|})$ to be a $(|\mathbb{D}_j|+1)$-dimensional binary  vector:   $d_i=1$ if and only if there exists $u_j\in {\cal M}_j(u^{j-1})$ such that  $\textsf{D}(u^j)$ is the $i$th element in $\mathbb{D}_j\cup\{\perp\}$, where $\mathbb{D}_j\cup\{\perp\}$ is sorted in any fixed order. Similarly, define $\mathbb{D}_{j-1}$ to be the set  of $\textsf{D}(u^{j-1})$ over all $u^{j-1}.$  Continuing the iterative definition till $\textsf{D}(u^1)$ and $\mathbb{D}_1$ is defined.  Let $D_j=|\mathbb{D}_j|$ for each $j$.  From our definition,  $D_j\le 2^{1+D_{j+1}}, \forall j$.

\vspace{.05in} \noindent{\bf Claim 2. } \quad {\em If $\textsf{\em D}(u^{j-1})=\textsf{\em D}(\bar{u}^{j-1})$ for some $u^{j-1}$ and $\bar{u}^{j-1}$, then (i) $\perp\in {\cal M}_j({u^{j-1}})$ if and only if $\perp\in {\cal M}_j({\bar{u}^{j-1}})$; (ii)  $u_j\in {\cal M}_j({u^{j-1}})\backslash\{\perp\}$ if and only if  there exists $\bar{u}_j\in {\cal M}_j(\bar{u}^{j-1})\backslash\{\perp\}$ such that  $\textsf{\em D}(u^{j})=\textsf{\em D}(\bar{u}^{j})$. }

\vspace{.05in} \noindent{\em Proof. } Let $\textsf{D}(u^{j-1})=\textsf{D}(\bar{u}^{j-1})=(d_0, d_1, \cdots, d_Q).$ W.L.O.G., $\perp$ is the 0th element in $\mathbb{D}_j\cup\{\perp\}$. Then, the claim follows from the definition:  (i) $d_0=1$ iff $\perp\in {\cal M}_j(u^{j-1})$ and $\perp\in {\cal M}_j(\bar{u}^{j-1})$;  (ii) $d_i=1$ for $i>0$ if and only if there exists $u_j\in {\cal M}_j(u^{j-1})$ such that  $\textsf{D}(u^j)$ is  the $i$th element in $\mathbb{D}_j\cup\{\perp\}$.  Especially, under the existence for (ii), $\textsf{D}(u^{j})=\textsf{D}(\bar{u}^{j})$ is the $i$th element in $\mathbb{D}_j\cup\{\perp\}$.  $\hfill\square$

\vspace{.10in}  Now we claim $|{\cal S}|\le D_1$;  otherwise, we construct an  Oscar who breaks  the authentication property as follows. Since $|{\cal S}|>D_1$, there must exist distinct $u_1, \bar{u}_1\in {\cal S}$ such that $\textsf{D}(u_1)=\textsf{D}(\bar{u}_1).$  Then, the code of Oscar is as follows.
\begin{itemize}

\item   Oscar provides $u_1$ to Alice as her source state input. When Alice  sends $u_1$ to Bob noiselessly, Oscar revises it to $\bar{u}_1$ and sends it to Bob.

\item      Assume the ($j$-1)th flow has been  handled and $\textsf{D}(u^{j-1})=\textsf{D}(\bar{u}^{j-1}).$ We handle  the $j$th flow for $j<L$ as follows.

\begin{itemize}
\item If Alice sends $u_j$ to Bob ($u_j\in {\cal M}_j({u^{j-1}})\backslash\{\perp\}$),  then  by Claim 2 there exists $\bar{u}_j\in {\cal M}_j({\bar{u}^{j-1}})\backslash\{\perp\}$ such that $\textsf{D}(u^{j})=\textsf{D}(\bar{u}^{j})$.  Oscar revises $u_j$ to $\bar{u}_j$ and sends it to Bob.
\item If Alice rejects with a local output  $u_j=\perp$, then $\perp\in {\cal M}_j(u^{j-1})$ by definition. By Claim 2, $\perp\in {\cal M}_j({\bar{u}^{j-1}})$, Oscar rejects Bob with a local output  $\bar{u}_j=\perp$.
\remove{\vspace{.05in} If Alice rejects (${\cal M}_j(u^{j-1})=\emptyset$), then Oscar  {\em rejects}  Bob (it is consistent as ${\cal M}_j(\bar{u}^{j-1})=\emptyset$ by Claim 2).}

 \item The case that Bob sends  $\bar{u}_j$ is  handled similarly.
\end{itemize}
\item Finally, when Alice outputs  $u_L=\perp$, the case is similar to $u_j=\perp$ for $j<L$; when Alice sends $u_L\in {\cal M}_L(u^{L-1})\backslash\{\perp\}$ to Bob, Oscar can not change it (in this case, we  define $\bar{u}_L=u_L$). However, based on the definition of ${\cal M}_L(u^{L-1})$ and  the previous item that $\textsf{D}(u^{L-1})=\textsf{D}(\bar{u}^{L-1})$, we know that $\bar{u}_L\in {\cal M}_L(u^{L-1})={\cal M}_L(\bar{u}^{L-1}).$   When Bob receives $\bar{u}_L$, if he outputs $\bar{u}_1$, then Oscar succeeds; otherwise, he fails.
\end{itemize}

 Now we analyze the success probability $p$ of Oscar. First of all, Alice is a sender with a uniformly random tape and especially is admissible. Thus, $u_j\in {\cal M}_j(u^{j-1})$ for any $j$. By our  analysis in the attack, $\bar{u}_j\in {\cal M}_{j}(\bar{u}^{j-1})$ as well. Thus, by the definition of {\em admissible} and the definition of ${\cal M}_L(\cdot)$,  Alice$'$ is an admissible sender in the execution (Alice$'$, Bob). By correctness,  Bob will output $\bar{u}_1$ with probability at least $1-\eta>\epsilon$, contradiction to the authentication property (as $\bar{u}_1\ne u_1$). Thus, $|{\cal S}|\le D_1.$  Finally, as $\log D_j\le 1+{D_{j+1}}$ for any $j$ (let $D_L=|{\cal X}^n$),  Claim 1 implies  that  $\log^{(L)} D_1\le 1+\log D_{L}=1+n\log{|{\cal X}|}$. Hence, $\log^{(L)} |{\cal S}| \le  1+n\log |{\cal X}|.$ Thus,
$\log^*|{\cal S}|=L+\log^*(\log^{(L)} |{\cal S}|)\le L+\log^*(1+n\log |{\cal X}|).$ This concludes the theorem for $\pi$ satisfying  the restrictions at the beginning.

For the general case, notice that for any $L$-round $\epsilon$ authentication protocol $\pi$, by Lemma \ref{le: finalW1} and Lemma \ref{le: startM}, there exists an $(L+4)$-round $(\epsilon+2^{-\beta n'})$-secure authentication protocol $\pi'$ with $n'=\gamma n$ for some constants $\beta>0, \gamma>0$ that satisfies the restriction at the beginning, where $n$ and $n'$ are respectively the number of channel $W_1$ uses in $\pi$ and $\pi'.$ Applying the above proof to $\pi'$, we conclude that $\log^*|{\cal S}|\le L+4+\log^*(1+n\gamma \log |{\cal X}|)\le L+4+\log^*(2^n)$ when $n$ large enough.   Hence, the theorem follows.   $\hfill\blacksquare$

\section{Lower bound on the success probability}
In this paper, we regard the DMC $W_1$ as an important resource and hope to  minimize the use of it.  For a fixed total length of messages over it and a fixed authentication error $\epsilon$, we might wish to authenticate a source space as large as possible. However, the following theorem shows that $\epsilon$ is very dependent on the message space on DMC $W_1$.

Our  idea is to present an Oscar that achieves a certain success probability. Roughly, when Alice is authenticating $S$ to Bob, Oscar blocks the communication between Alice and Bob. In addition, Oscar plays the role of `Bob' to interact with Alice. At the same time, Oscar starts an independent session to play the role of `Alice' to authenticate a new message $S'$ to Bob, except that he uses Alice's DMC messages in the previous session as his own. Here two authentication sessions are independent, except that they use the same DMC messages. By calculation, we can show that two independent sessions share the same DMC messages with probability at least $2^{-H(F)}$. When this event occurs, Bob will accept $S'$, except a completeness error error. So Oscar succeeds with probability at least $2^{-H(F)}-\delta-\frac{1}{|{\cal S}|}$, where $\frac{1}{|{\cal S}|}$ accounts for the possibility of  $S=S'$.  The formal detail is as follows.

\begin{theorem} Let $\pi$ be an $\epsilon$-secure authentication protocol in our model for source space ${\cal S}$ with correctness  error $\delta$.  Assume $F$ is the concatenation of messages over DMC $W_1$ by  Alice (if some flow does not contain a DMC message, use an empty symbol to represent  the DMC message in this flow).  Let ${\cal F}$ be  the space of $F$.  Then, $\epsilon\ge 2^{-H(F)}-\delta-\frac{1}{|{\cal S}|}$.
Especially, $\epsilon\ge \frac{1}{|{\cal F}|}-\delta-\frac{1}{|{\cal S}|}$.
\end{theorem}
\noindent{\bf Proof. } We now present   a strategy for  Oscar to achieve the claimed lower bound.   Oscar  first generates $S'\leftarrow {\cal S}$ and then simulates  two parties: Alice$'$ and  Bob$'$ to conduct a type I  attack (denoted by $\Gamma$) as follows.
\begin{itemize}
\item When Alice interacts with Bob for authenticating $S\leftarrow {\cal S}$, Bob$'$ intercepts and blocks all the messages from Alice, except the messages over DMC $W_1.$ In addition, Bob$'$, in the role of Bob,  interacts with Alice faithfully, except that he simulates the output of $W_1$ using the input from Alice  (recall that Oscar can see the input of Alice over $W_1$).  In addition,  Alice$'$ intercepts and blocks all the messages from Bob. She then interacts with Bob faithfully to authenticate $S'$, except that she regards each message over DMC $W_1$ from Alice   as her own message to Bob.
\end{itemize}
In this attack,  Oscar succeeds if and only if Bob outputs $S'$ (denoted by event \textsf{Good}) and $S'\ne S.$  So
$P(succ(\mbox{Oscar})\ge P(\textsf{Good})-P(S'=S)= P(\textsf{Good})-{1}/{|{\cal S}|}.$

In the following, we analyze $P(\textsf{Good})$. Toward this, we consider a mental variant (denoted by $\Gamma'$) of Oscar's attack $\Gamma$, where the difference is  as follows.
\begin{itemize}
\item[-] Bob$'$ does not use the simulated output of $W_1$ and instead he can also intercept and block the channel $W_1$ and use the channel output.
\item[-] Alice$'$ does not use the messages $W_1$ from Alice as her own $W_1$ messages to Bob. Instead, she can send messages directly onto $W_1$  and Bob can receive the corresponding output.
\end{itemize}
In other words,  Bob$'$ and Alice$'$ is changed such that (Alice, Bob$'$) and  (Alice$'$, Bob) maintain two independent protocol executions, where the former is to authenticate $S\leftarrow {\cal S}$ while the latter is to authenticate $S'\leftarrow {\cal S}.$

Let $F_1$ be the messages over $W_1$ in execution (Alice, Bob$'$) and $F_2$ be the messages over $W_1$ in execution (Alice$'$, Bob).  Observe that a simulated $W_1$ and a  real $W_1$ have the same statistical characteristics. It follows that, conditional on $F_1=F_2$, $\Gamma'$ and $\Gamma$ are distributed identically. Let $P^\Gamma(\textsf{E})$ denote the event $\textsf{E}$ in an experiment $\Gamma.$  Then,
\begin{align}
\nonumber
P^\Gamma(\textsf{Good})\ge&  P^\Gamma(\textsf{Good}| F_1=F_2) P^{\Gamma}(F_1=F_2)\\
\nonumber
=& P^{\Gamma'}(\textsf{Good}| F_1=F_2) P^{\Gamma}(F_1=F_2)\\
\nonumber
\ge & P^{\Gamma'}(\textsf{Good}| F_1=F_2) P^{\Gamma'}(F_1=F_2)\\
\nonumber
& (\mbox{$P^{\Gamma}(F_1=F_2)=1$ by definition of $\Gamma$})\\
=& P^{\Gamma'}(\textsf{Good}, F_1=F_2) \label{eq: boundp}
\end{align}
Further, in $\Gamma'$,  executions (Alice, Bob$'$) and (Alice$'$, Bob) are independent. Also, $F_1$ is an event in the execution of (Alice, Bob$'$) while $(\textsf{Good}, F_2)$ is an event in the execution of (Alice$'$, Bob).
 So $F_1$ is independent of  $(\textsf{Good}, F_2)$. Thus,

\quad Eq. (\ref{eq: boundp})

$= \sum_{a\in {\cal F}} P^{\Gamma'}(\textsf{Good}, F_2=a)P^{\Gamma'}(F_1=a)$

$\ge \sum_{a\in {\cal F}} P^{\Gamma'}(F_2=a)P^{\Gamma'}(F_1=a)-\delta$

\quad $/*$ execution (Alice$'$, Bob) is faithfully according to $\pi$

\quad and so $P^{\Gamma'}(\textsf{Good})\ge 1-\delta$.  $*/$

$= \sum_{a\in {\cal F}} P^2_F(a)-\delta, $

\quad  $/*$ $F_1, F_2$ are i.i.d.  according to  the corresponding RV $F$

\quad of a faithful execution of $\pi$. $*/$

$\ge 2^{-H(F)}-\delta, $

\quad $/*$ {$\log(\sum_xP_X^2(x))\ge -H(X)$ as $\log(x)$ is concave} $*/$

\vspace{.05in} \noindent This gives the first conclusion. The second one follows from $H(F)\le {|{\cal F}|}.$ This completes the proof. $\hfill\blacksquare$

\section{The Capacity of Non-interactive Authentication over any DMC}
In this section, we study  a non-interactive case of the keyless authentication in our model: the protocol consists only of one message flow  $(X^{n}, u)$ sent from Alice to Bob, where $X^n$ is over channel $W_1$ and $u$ is over the noiseless channel.  The authentication capacity in this setting with BSCs $W_1$ and $W_2$  was  obtained in \cite{Jiang14}. In the following, we extend it  to the general DMC setting.

Our idea is as follows. By Lemma \ref{le: Vd}, we have a subset ${\cal C}$ of size $|{\cal X}^{n(1-\delta)}$ for an arbitrarily  small $\delta>0$ such that any two elements in ${\cal C}$ has a large distance.  By Lemma \ref{le: cap}, if we send $C_i\in {\cal C}$ over DMC, Bob will not confuse it with $C_j\in {\cal C},$ in the sense of the presence of a type I attack.  So ${\cal C}$ can be used to authenticate a source space of size $|{\cal X}|^{n(1-\delta)}$ against type I attack. A type II attack can be combated using the same idea in SetAuth$^*$. This gives a scheme with an authentication rate of $(1-\delta)\log|{\cal X}|$. Since $\delta$ is arbitrarily small, any rate less than $\log|{\cal X}|$ can be achieved. On the other hand, it is obvious that the rate can not surpass $\log |{\cal X}|$ as the noiseless channel is insecure and hence one codeword over DMC $W_1$ can authenticate at most one source state.

\begin{theorem} The capacity of  a non-interactive authentication in our model with $W_1$  non-redundant and $Cov(W_1)\not\subseteq Cov(W_2)$   is $\log|{\cal X}|$.
\end{theorem}
\noindent{\bf Proof. } {\em Achievability. } For any $\alpha\in (1/n, 1/2]$, by Lemma \ref{le: Vd}, there exists  ${\cal C}\subseteq{\cal X}^n$ such that any two elements in it have  distance at least $\alpha n$ and that   $|{\cal C}|\ge \frac{|{\cal X}|^{n(1-\alpha-\frac{h(\alpha)}{\log|{\cal X}|})}}{\alpha n}. $  Now let  ${\cal C}=\{C_1, \cdots, C_N\}$.

   Let $k=\sqrt{n}$.  Since $Cov(W_1)\not\subseteq Cov(W_2)$,  there exists $a\in {\cal X}$ such that  $W_1(\cdot |a)\not\in Cov(W_2)$.
   So $\Delta(W_1(\cdot|a), Cov(W_2))=\xi$ for some $\xi>0$.   Let $\epsilon=\min\{\frac{\xi}{4|{\cal Z}|}, \frac{\alpha\Theta}{2}\}$ and $\epsilon'=2e^{-\frac{k\xi^2}{8|{\cal Z}|^2}}, $ where $\Theta$ is defined in Lemma \ref{le: cap} for the non-redundant DMC $W_1$.   We construct the protocol for Alice to  authenticate  $s\in [N]$ as follows.
 \begin{itemize}
 \item[1.]  Alice sends $a^k|C_s$ over channel $W_1$ and $s$ over the noiseless channel.
  \item[2.] Upon $Z^{n+k}$ from channel $W_1$ and $s'$ from the noiseless channel, Bob   checks if   ${Z^k}\in \textsf{T}^k_{[W_1]_\epsilon}(a^k)$ and  $Z_{k+1}^{k+n}\in \textsf{T}_{[W_1]_{\epsilon}}^n(C_{s'})$. If yes, he outputs $s'$; otherwise, he rejects.
 \end{itemize}

Consider a type II attack first. Assume Oscar sends ${X}^{k+n}$ over $W_2$. We claim that  $P_{Z^k}(\textsf{T}^k_{[W_1]_\epsilon}(a^k))\le \epsilon'$ (in other words, $P_{Z^k}\Big{(}|T_{Z^k}(u)-W_1(u|a)|\le \frac{\epsilon}{|{\cal Z}|}, \mbox{for all $u\in {\cal Z}$}\Big{)}\le \epsilon'$).
Otherwise, by Lemma \ref{le: neg},
\begin{align*}
 \Delta\Big{(}W_1(\cdot|a),  Cov(W_2)\Big{)}\le & |{\cal Z}|\epsilon+|{\cal Z}|\sqrt{\frac{\ln{(2/\epsilon')}}{2k}}\\
 \le & \frac{\xi}{4}+\frac{\xi}{4}<\xi,
\end{align*}
 which contradicts   $\Delta(W_1(\cdot|a), Cov(W_2))=\xi.$ Thus, a type II attack succeeds with probability at most $\epsilon'=2e^{-\frac{k\xi^2}{8|{\cal Z}|^2}}.$

We now consider a type I attack.   In this case, Oscar  succeeds only if $Z_{k+1}^{k+n}\in \textsf{T}^n_{[W_1]_{\epsilon}}(C_{s'})$  for $s'\ne s.$ However,   $d_H(C_s, C_{s'})>\alpha n.$ By Lemma \ref{le: cap},
 \begin{align}
W_1(\textsf{T}^n_{[W_1]_{\epsilon}}(C_{s'})|C_s)\le & 2^{-\frac{2n(\alpha\Theta-\epsilon)^2}{|{\cal X}|^2|{\cal Z}|^2}}\le 2^{-\frac{n\alpha^2\Theta^2}{2|{\cal X}|^2|{\cal Z}|^2}},
 \end{align}
exponentially small!

Authentication rate is  $
\lim_{n\rightarrow \infty}\frac{1}{n+k}\log \frac{|{\cal X}|^{n(1-\alpha-\frac{h(\alpha)}{\log|{\cal X}|})}}{\alpha n}=[1-\alpha-\frac{h(\alpha)}{\log|{\cal X}|}]\log|{\cal X}|.$  Since $\alpha$ is arbitrarily  small, any rate less than $\log|{\cal X}|$ can be achieved.

\noindent{\em Converse. } Since any point in ${\cal X}^n$ can be a codeword for at most one source $s$ (recall the noiseless channel can be modified arbitrarily), the authentication rate is at most $\log|{\cal X}|$.   $ \hfill\blacksquare$

\section{Conclusion}
In this paper, we further studied   the keyless authentication problem in the  noisy model of our previous work \cite{Jiang14}. We extended the  construction in \cite{Jiang14}. If the message space is ${\cal S}$ and the number of channel $W_1$ uses is $n$, then our new protocol has a round complexity $\log^*|{\cal S}|-\log^* n+4$. Here $n$ can be chosen independent of ${\cal S}$ while this is impossible in the protocol of \cite{Jiang14}. We proved a lower bound $\log^*|{\cal S}|-\log^*n-5$ on the round complexity. We also obtained a lower bound on the success probability. Finally, we showed the capacity for a non-interactive authentication under  general DMCs $W_1, W_2$  is log $|{\cal X}|$,  which extends the result under   BSCs in \cite{Jiang14}.

\remove{ \section*{Acknowledgments}
 \noindent This work  was supported by National 973 Program of China (No. 2013CB834203).
}


\begin{thebibliography}{00}

\bibitem{AC93} R. Ahlswede, I. Csisz\'{a}r, ``Common randomness in information theory and cryptography. Part I: secret sharing'', {\em IEEE Transactions
on Information Theory}, vol. 39, pp. 1121-1132, 1993.

\bibitem{HS11} H. Ahmadi, R. Safavi-Naini, ``Secret Keys from Channel Noise'', in {\em Proc.  Adavances in Cryptology-EUROCRYPT 2011}, K. G.  Paterson (Ed.), LNCS 6632, pp. 266-283, 2011.


\bibitem{BLT12} P. Baracca, N. Laurenti, and S. Tomasin, ``Physical Layer Authentication over
MIMO Fading Wiretap Channels'', {\em IEEE Transactions on Wireless Communications},  vol. 11, no. 7, pp. 2564-2573, July 2012.

\bibitem{BINS06} J. Barros, H. Imai, A. Nascimento, S. Skludarek,  ``Bit commitment over Gaussian channels'', {\em in Proc. IEEE International
Symposium on Information Theory  2006}, pp. 1437-1441, 2006.


\bibitem{BCK98}  M. Bellare, R. Canetti, and H. Krawczyk, a modular
approach to the design and analysis of authentication and key
exchange protocols, {\em  Proceedings of the Thirtieth Annual ACM
Symposium on the Theory of Computing}, pp. 419-428, 1998, Dallas,
Texas, USA.

\bibitem{BTV12} M. Bellare, S. Tessaro, and A. Vardy, ``Semantic security for the wiretap channel'',
in {\em Adavances in Cryptology} (Lecture Notes in Computer Science),
vol. 7417, R. Safavi-Naini and R. Canetti, Eds. Berlin, Germany: Springer-Verlag, 2012,
pp. 294-311.


\bibitem{BB11} M. Bloch, J.  Barros, {\em Physical Layer Security:  From Information Theory to Security Engineering}, Cambridge University Press, 2011.

\bibitem{BBM07} M. Bloch, J.  Barros, S.  McLaughlin, ``Practical information-theoretic commitment'',  in {\em Proc. Allerton Conference Communication,
Control, and Computing 2007}, pp. 1035-1039, 2007.


\bibitem{CK88} C. Cr\'{e}peau and J. Kilian, ``Achieving oblivious transfer using weakened security assumptions'', in {\em Proc. 29th Annual Symposium on Foundations of Computer Science (FOCS'88)}, pp. 42-52, 1988.

\bibitem{Crea97}  C. Cr\'{e}peau,   ``Efficient Cryptographic Protocols Based on Noisy Channels'', in {\em Proc. Advances in Cryptology-EUROCRYPT 1997}, J. Borst et al. (Eds.), LNCS 1233, pp. 306-317, 1997.

\bibitem{CMW04} C. Cr\'{e}peau, K. Morozov, S. Wolf, ``Efficient unconditional oblivious transfer from almost any noisy channel'', in {\em Proc. Security in Communication Networks 2004}, C. Cr\'{e}peau (Ed.), LNCS 3352, pp. 47-59, 2004.

\bibitem{CK78} I. Csisz\'{a}r and J. K\"{o}rner, Broadcast channels with confidential messages, {\em IEEE Transactions on Information Theory}, Vol. IT-24, No. 3, May 1978, pages 339-348.

\bibitem{CK11} I. Csisz\'{a}r and J. K\"{o}rner, {\em Information Theory: Coding Theorem for Discrete Memoryless System,}  Combridge University Press, 2011.


\bibitem{CN00} I. Csisz\'{a}r and P. Narayan,  ``Common randomness and secret key generation with a helper'', {\em IEEE Transactions on  Information Theory}, vol. 46, pp. 344-366, 2000.



\bibitem{GMS74} E. N. Gilbert, F. J. MacWilliams and N. J. Sloane, ``Codes which detect deception'', {\em Bell System Technical Journal}, Vol 53, No. 3, pp. 405-424, 1974.


\bibitem{HP85} D. R. Hughes and F. C. Piper, {\em Design Theory},  Cambridge University Press, 1985.

\bibitem{Jiang14} S. Jiang, Keyless Authentication in  a Noisy Model, {\em IEEE Transactions on Information Forensics and Security}, Vol. 9, No. 6, pp. 1024-1033, 2014.

\bibitem{Jiang14a} S. Jiang, (Im)possibility of Deterministic  Commitment over a Discrete Memoryless Channel, {\em IEEE Transactions on Information Forensics and Security}, Vol. 9, No. 9, pp. 1406-1415, 2014.


\bibitem{KDW08} A. Khisti, S. Diggavi, G. Wornell, ``Secret key generation with correlated sources and noisy channels'', in  {\em Proc. IEEE International
Symposium on Information Theory 2008}, pp. 1005-1009 (2008).

\bibitem{KY+07} V. Korzhik, V. Yakovlev, G. M. Luna, R. Chesnokov, ``Performance Evaluation of Keyless Authentication Based on Noisy Channel'',
In {\em Proc. MMM-ACNS 2007},  V. Gorodetsky et al.  (Eds.), Springer-Verlag, Berlin,  pp. 115-126, 2007.




\bibitem{LGP09} L. Lai, H. ElGamal and H. V. Poor, ``Authentication
over noisy channels'', {\em IEEE Trans. on Inf. Theory}, vol.
55, no. 2, pp. 906-916, Feb. 2009.


\bibitem{LPS08} Y. Liang, H. V. Poor, S. Shamai,
``Information Theoretic Security'', {\em Foundations and Trends in Communications and Information Theory}, vol 5, nos 4-5, pp 355-580, Now Publishers, Hanover, MA, USA, 2008.






\bibitem{Mau93} U. Maurer, ``Secret key agreement by public discussion from common information'', {\em IEEE Transaction on Information Theory},
vol. 39, pp. 733-742, 1993.


\bibitem{Mau03a} U. Maurer and S. Wolf,  ``Secret-key agreement over unauthenticated public channels - part I: definitions and a completeness'',
{\em IEEE Transactions on Information Theory}, vol. 49, pp. 822-831 (2003).


\bibitem{MWC05} E. Martinian, G.W. Wornell, and B. Chen,
``Authentication with distortion criteria'', {\em IEEE Transactions on
Information  Theory}, vol. 51, pp. 2523-2542, 2005.

\bibitem{NW06} A. Nascimento and A.  Winter, ``On the oblivious transfer capacity of noisy correlations'', in  {\em Proc. IEEE International Symposium
on Information Theory 2006}, pp. 1871-1875, 2006.


\bibitem{RSA78} R. Rivest, A. Shamir and L. Adleman, ``A
method for obtaining digital signatures and public-key
cryptosystems'', {\em Communications of ACM}, vol. 2, pp. 120-126,  February 1978.




\bibitem{YBS08} P. L. Yu, J. S. Baras, and B. M. Sadler, “Physical-layer authentication,”
{\em IEEE Trans. Inf. Forensics and Security}, vol. 3, no. 1, pp. 38-51, Mar.
2008.


\bibitem{WNI03} A. Winter, A. Nascimento, and H. Imai, ``Commitment capacity of
discrete memoryless channels,” in {\em Proc. 9th IMA Conf. Coding and
Cryptography (WCC 2003)}, K.G. Paterson (Ed.), LNCS 2898, pp. 35-51, 2003.

\bibitem{Wyner75}  A. D. Wyner, ``The wire-tap channel'',  {\em Bell System Technical Journal}, vol. 54, pp. 1355-1367, 1975.

\end{thebibliography}
\end{document}